\documentclass[showpacs,preprintnumbers, amsmath,amssymb,superscriptaddress,preprint,floatfix]{revtex4}
\usepackage{graphicx}
\draft
\begin{document}
\newcommand {\bb}{\bibitem}
\newcommand {\be}{\begin{equation}}
\newcommand {\ee}{\end{equation}}
\newcommand {\bea}{\begin{eqnarray}}
\newcommand {\eea}{\end{eqnarray}}
\newcommand {\nn}{\nonumber}

\title{Perspectives on Nodal Superconductors}

\author{Kazumi Maki}
\author{Stephan Haas}
\author{David Parker}

\address{Department of Physics and Astronomy, University of Southern
California, Los Angeles, CA 90089-0484 USA}

\author{Hyekyung Won}

\address{Department of Physics, Hallym University,
Chuncheon 200-702, South Korea}

\date{\today}

\begin{abstract}
In the last few years the gap symmetries of many new superconductors,including Sr$_2$RuO$_4$,
CeCoIn$_5$, $\kappa$-(ET)$_{2}$Cu(NCS)$_{2}$, YNi$_{2}$B$_{2}$C and PrOs$_{4}$Sb$_{12}$, have been
identified via angle-dependent magnetothermal conductivity measurements.
However, a controversy still persists as to the nature of the superconductivity in Sr$_2$RuO$_4$. 
For PrOs$_{4}$Sb$_{12}$, spin-triplet superconductivity has recently been
proposed.  Here, we also propose g-wave superconductivity for UPd$_2$Al$_3$
(i.e., $\Delta({\bf k})=\Delta\cos(2\chi), \chi = ck_{z}$) based on recent
thermal conductivity data.

\end{abstract}
\maketitle
 
\noindent{\it \bf 1.  Introduction}

After the appearance of heavy-fermion superconductors and organic superconductors 
in 1979 the gap symmetries of these new compounds have been a central issue\cite{1}.  However,
until recently only the d$_{x^{2}-y^{2}}$-wave symmetry of the gap function $\Delta({\bf k})$ 
in high-T$_c$ cuprates has been established by the elegant Josephson 
interferometry\cite{2} and the angle resolved photoemission spectra (ARPES)\cite{3}. Unfortunately, so far
these powerful techniques are unavailable for heavy-fermion superconductors and organic
superconductors with lower superconducting transition temperatures T$_c \leq 10 K$.

In the last few years, Izawa et al have established the gap symmetries of superconductivity in
Sr$_2$RuO$_4$\cite{4},CeCoIn$_5$\cite{5}, $\kappa$-ET$_{2}$Cu(NCS)$_{2}$\cite{6}, 
YNi$_{2}$B$_{2}$C\cite{7}
and PrOs$_{4}$Sb$_{12}$\cite{8} through the angle dependent magnetothermal conductivity.  This
breakthrough relies in part on the availability of high-quality single crystals of these
compounds and in part on the theoretical development initiated by Volovik.\cite{9}  Last year,
we have reviewed the progress in \cite{10}.

In the present paper, we focus on 3 recent topics in nodal superconductors.  In spite of
ample evidence for f-wave superconductivity in Sr$_{2}$RuO$_4$ \cite{10} the controversy
regarding this compound appears to continue.  Therefore in section 2 we discuss the 
angle dependent magnetospecific
heat data by Deguchi et al \cite{11}.  Now evidence for spin-triplet superconductivity
in PrOs$_{4}$Sb$_{12}$ is mounting.  In section 3, we describe p+h-wave superconductivity
for the A and B phases in PrOs$_4$Sb$_{12}$\cite{12}.  Recently angle-dependent thermal 
conductivity data
in the vortex state in UPd$_2$Al$_3$ has been reported.\cite{13} In section 4 we 
analyze the angle-dependent
magnetothermal conductivity $\kappa_{yy}$ when the field is rotated within the z-x plane, and we
conclude that $\Delta({\bf k})$ in UPd$_2$Al$_3$ is given by $\Delta({\bf k}) 
= \Delta \cos(2\chi)$\cite{14}.
In Fig. 1 we show the new $|\Delta({\bf k})|'s$ so far identified.
\begin{figure}[h]
\includegraphics[width=3.5cm,height=4.5cm]{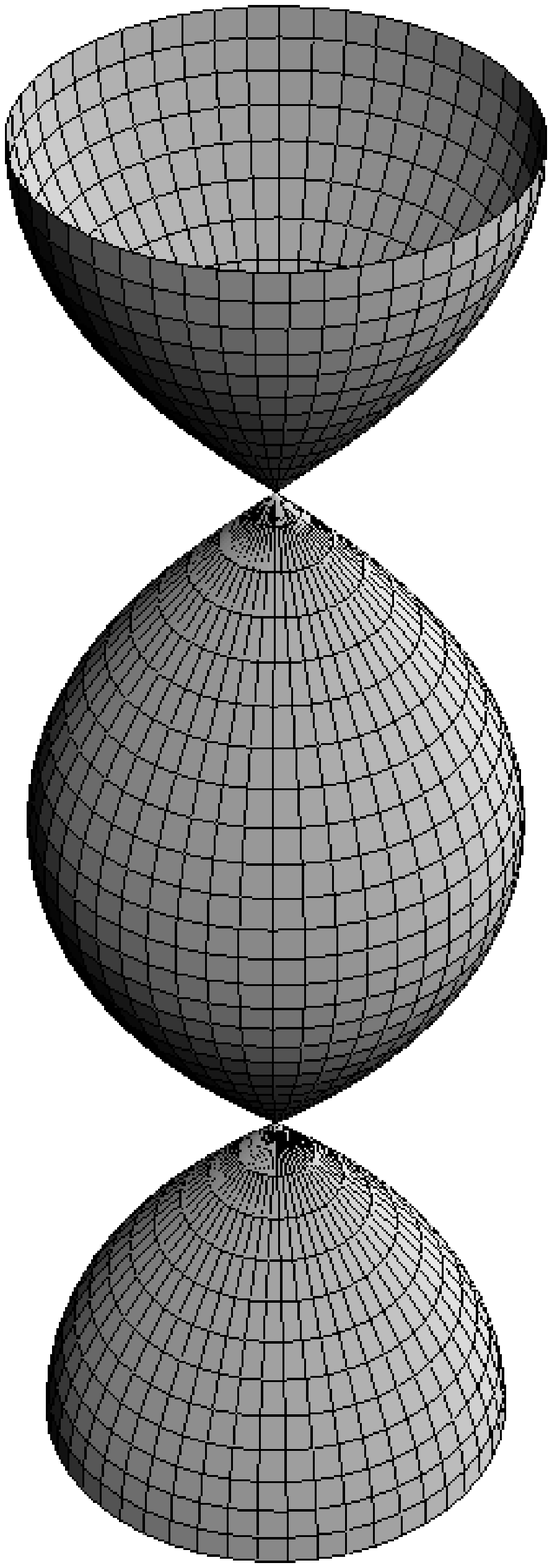}
\includegraphics[width=5.5cm]{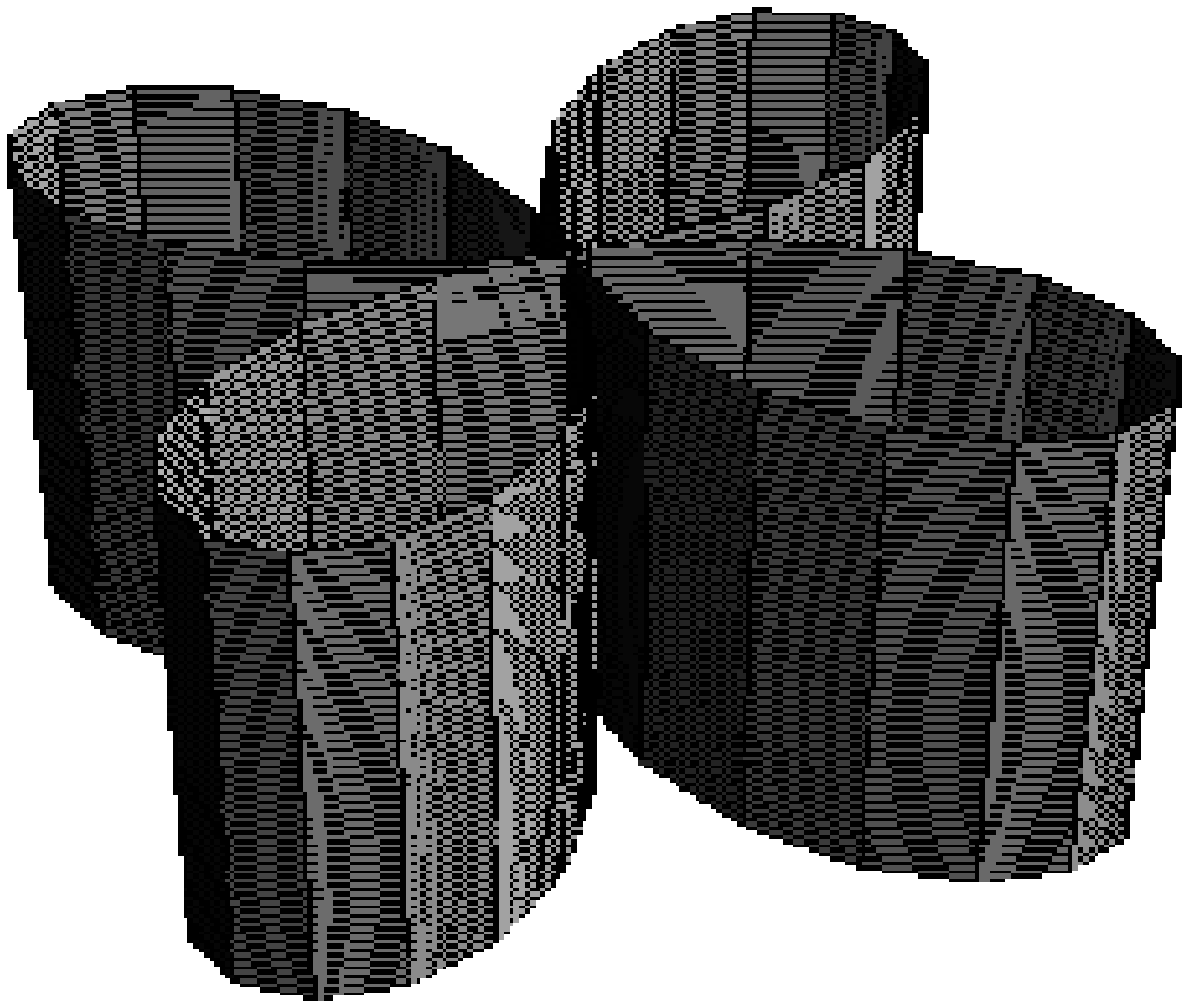}
\includegraphics[width=6.25cm]{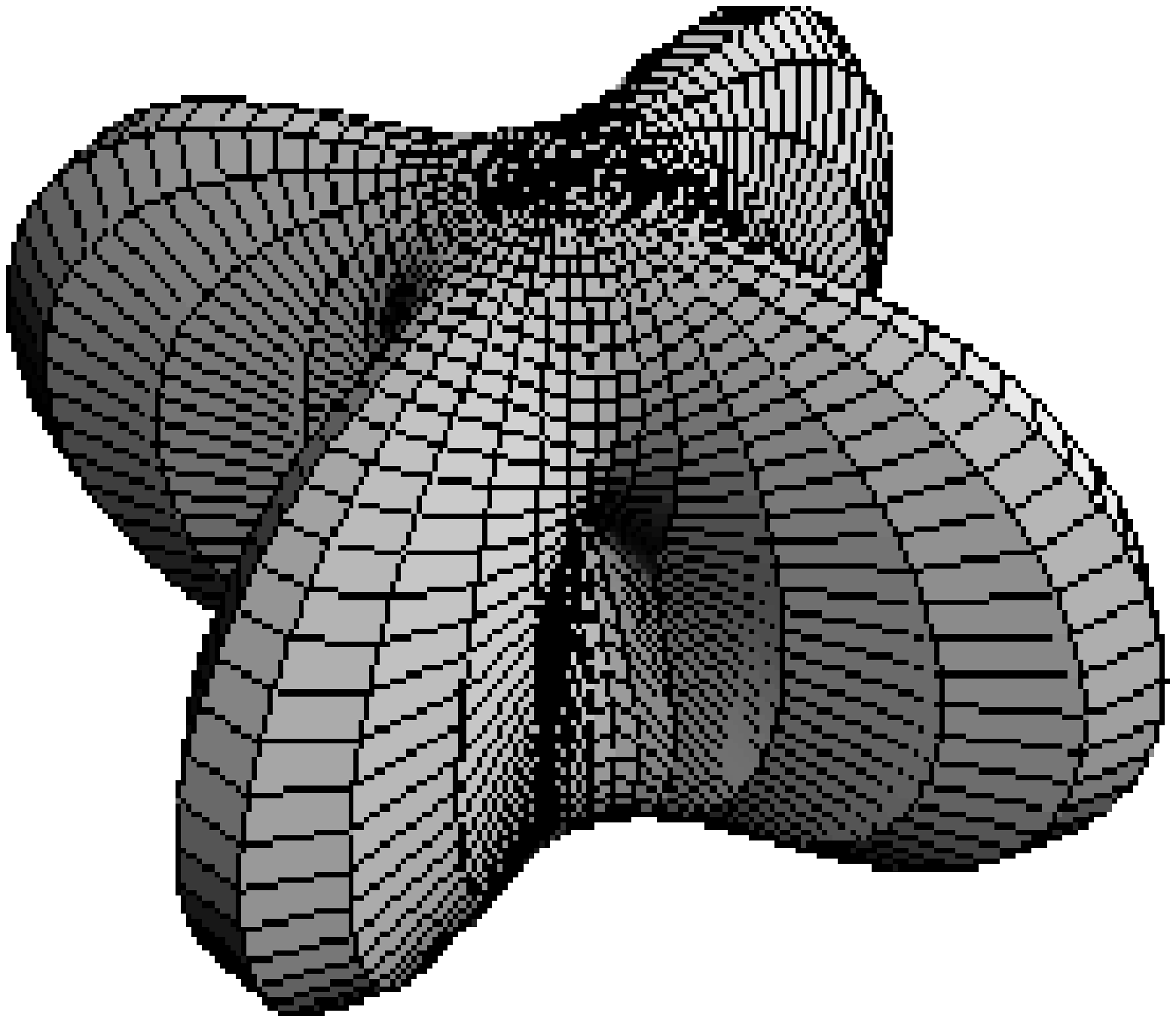}
\includegraphics[width=5cm]{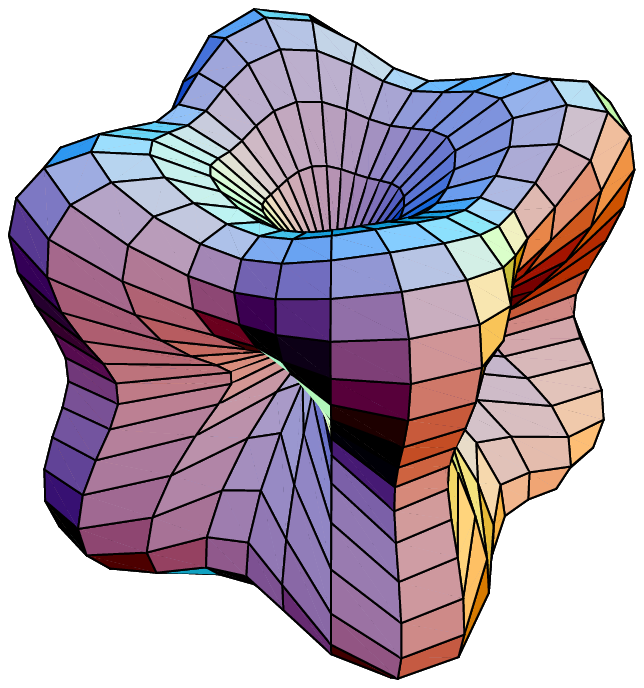}
\includegraphics[width=5cm]{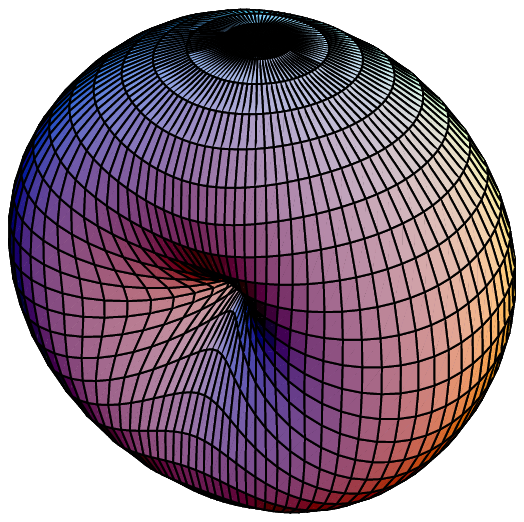}
\caption{From top left, 2D f-wave - Sr$_2$RuO$_4$, d$_{x^{2}-y^{2}}$-wave - CeCoIn$_{5}$ and 
$\kappa$-(ET)$_{2}$Cu(NCS)$_{2}$,s+g-wave - YNi$_{2}$B$_{2}$C, p+h-wave - PrOs$_4$Sb$_{12}$ - 
A phase, p+h-wave - PrOs$_4$Sb$_{12}$ - B phase.}
\end{figure}

\noindent{\it \bf 2. F-wave Superconductivity in Sr$_2$RuO$_4$}

Superconductivity in Sr$_2$RuO$_4$ was discovered in 1994\cite{15}. Sr$_2$RuO$_4$ is
an isocrystal to La$_2$CuO$_4$, but it is metallic down to low temperatures and becomes
superconducting around T = 1.5 K.  An early review on Sr$_2$RuO$_4$ can be found in Ref.\cite{16}.
From the analogy to superfluid $ ^{3}$He Rice and Sigrist\cite{17} proposed 2D p-wave
superconductivity.  Indeed spin-triplet pairing and related chiral symmetry-breaking have
been established \cite{18,19,20}.  As sample quality improved around 1999, both the specific heat
data \cite{21} and the superfluid density \cite{22} 
indicated nodal structure in the superconducting order parameter of
Sr$_2$RuO$_4$.  These findings ruled out  p-wave superconductivity and its generalization\cite{23}.  Therefore,
a variety of f-wave order parameters were suggested. \cite{24}  In Fig.2 and Fig.3 we show the
specific heat data \cite{21} and the superfluid density data\cite{22} 
compared with a variety of models.
\begin{figure}[h]
\includegraphics[width=6cm]{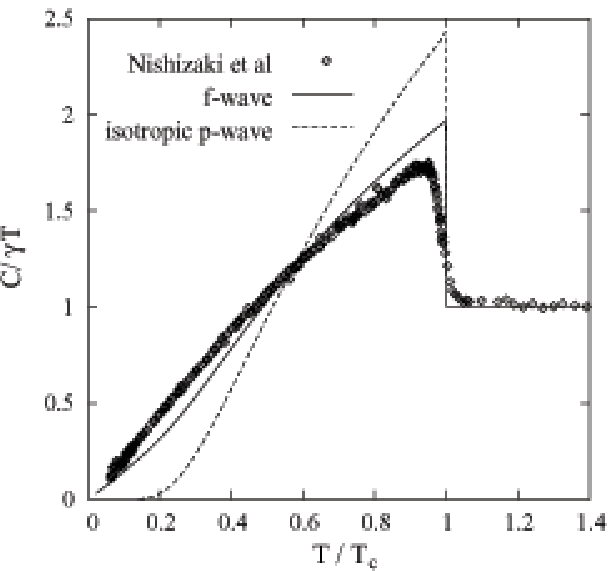}
\caption{Specific heat for 2D p-wave and f-wave models for Sr$_2$RuO$_4$.}
\end{figure}
\\
\begin{figure}[h]
\includegraphics[width=6cm]{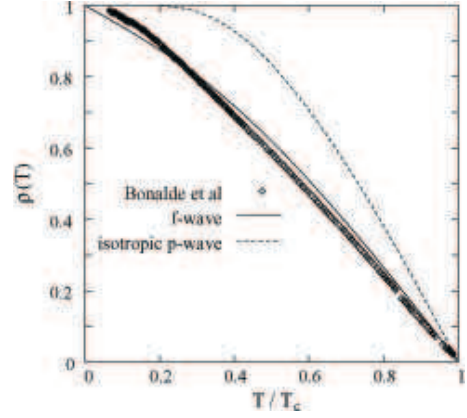}
\caption{Superfluid density for 2D p-wave and f-wave models for Sr$_2$RuO$_4$.}
\end{figure}
However, these experiments cannot tell us about the nodal structure of $\Delta({\bf k})$.
In a quasi-2D system such as Sr$_2$RuO$_4$, the line nodes in $\Delta({\bf k})$ can be either
vertical or horizontal.  But vertical nodes are  incompatible with the angular
dependent magnetothermal conductivity \cite{4} and the ultrasonic attenuation data \cite{25}.
Furthermore, Ref. 4 indicates that the horizontal nodes are far away from $\chi_{0}$=0.  This suggests
$\Delta({\bf k}) = {\bf d}e^{\pm i \phi} \cos(\chi)$, i.e. 2D f-wave superconductivity\cite{26}.

This interpretation is contested by Deguchi et al \cite{11}.  They measured the magnetospecific heat
of  Sr$_2$RuO$_4$ in a rotating magnetic field down to 100 mK and found cusp-like features
only in the regime $0.12 K < T < 0.31 K$.  From our earlier analysis of s+g-wave 
superconductivity\cite{27,28}, we deduce that there should be a point-like minigap with $\Delta_{min} \sim 
0.1 K$.  The simplest triplet gap function which has these minigaps is
\bea
\Delta({\bf k})= {\bf d}e^{\pm i \phi}(1 + a \cos(4\phi)\cos(\chi)) 
\eea
where $|1-a| \leq 0.1$.  Deguchi et al have proposed the Miyake-Nariyiko (MN) model \cite{29}, in
order to describe the measured specific heat.  However, it is easy to see that the MN model cannot
give the cusp-like features in the magnetospecific heat.  Also, the MN model cannot
describe the observed $T^{2}$ specific heat or the T-linear dependence of the superfluid density.
Moreover, the angular dependent thermal conductivity data and the universal heat conduction in
$\kappa_{xx}$ by Suzuki et al \cite{30} are incompatible with the MN model.  
Therefore further experiments on 
Sr$_2$RuO$_4$ are highly desirable.  We have proposed that the optical conductivity \cite{31},
the Raman scattering\cite{32} and the supercurrent \cite{33,34} in Sr$_2$RuO$_4$ will provide
further insight on its superconductivity.

\noindent{\it \bf 3. Triplet Superconductivity in PrOs$_4$Sb$_{12}$}

Superconductivity with T$_c$ = 1.8 K has
been discovered very recently in the skutterudite PrOs$_4$Sb$_{12}$\cite{35,36,37}.  
Angle-dependent thermal conductivity data on
this system has revealed a multi-phase structure, characterized by a gap function with
point nodes.\cite{8}  In order to account for this nodal structure s+g-wave superconductivity
has been proposed. \cite{10,38}

Recently there has been mounting experimental evidence for triplet superconductivity in this
compound.  First, from $\mu$SR measurements Aoki et al discovered a remnant magnetization in
the B-phase of this compound, indicating triplet pairing. \cite{39}  Second, the thermal
conductivity measurement in a magnetic field down to low-temperature ($T > 150$ mk) indicates
$\kappa_{zz} \sim T$ and H \cite{40}, consistent with triplet pairing.  Later we shall discuss
$\kappa_{zz}$ measured in a magnetic field rotated within the z-x plane.  This data is fully
consistent with triplet p+h-wave superconductivity in PrOs$_4$Sb$_{12}$.  Finally, a recently 
reported NMR result for the Knight shift by Tou et al \cite{41} also suggests the triplet pairing.
Here we propose p+h-wave order parameters
\bea
{\bf \Delta}_{A}({\bf k})&=&\frac{3}{2}{\bf d} e^{\pm i\phi_{1} \pm i \phi_{2} \pm i 
\phi_{3}}(1-\hat{k}_{1}^{4}-\hat{k}_{2}^{4}-\hat{k}_{3}^{4}))\\
{\bf \Delta}_{B}({\bf k})&=&{\bf d} e^{\pm i\phi_{3}}(1-\hat{k}_{3}^{4})
\eea
for the A-phase and B-phase of PrOs$_4$Sb$_{12}$, respectively, where $e^{\pm i \phi_{1}}=
\hat{k_{2}} \pm i \hat{k_{3}}$, etc.
These order parameters have nodal structures consistent with the angle
dependent thermal conductivity data \cite{8}, assuming that in the experiment
the nodes in the B-phase are aligned parallel to the y-axis.  We note $|\Delta({\bf k})|$
in the A-phase has cubic symmetry whereas in the B-phase it has axial symmetry.  Furthermore,
it appears that in the slow field-cooled situation the nodes in the B-phase are aligned parallel
to the magnetic field.  At least this is the simplest way to interpret the superfluid density
measurement by Chia et al \cite{42,43}.

Here we give expressions for the thermal conductivity $\kappa_{zz}$ in a 
magnetic field in the superclean limit ($\sqrt{\Gamma \Delta} \ll v\sqrt{eH}$),
\bea
\kappa_{zz}/\kappa_{n}&=& \frac{v^{2}eH}{8\Delta^{2}}\sin^{2}(\theta)\,\,\,\,\,\, ,A-phase \\
&=& \frac{3v^{2}eH}{64\Delta^{2}}\sin^{2}(\theta)  \,\,\,\,\, ,B-phase
\eea
and in the clean limit ($v\sqrt{eH} \ll \sqrt{\Gamma \Delta}$)
\bea
\kappa_{zz}/\kappa_{00} &=& 1 + \frac{3v^{2}eH}{40\Gamma \Delta} \ln(\sqrt{\frac{2\Delta}{\Gamma}}) 
\sin^{2}(\theta) \ln(\frac{\Delta}{v\sqrt{eH}\sin(\theta)}),\,\,\,\,\,\, A-phase \\
&=& 1+\frac{v^{2}eH}{12\Gamma \Delta} \ln(\sqrt{\frac{2\Delta}{\Gamma}}) 
\sin^{2}(\theta) \ln(\frac{\Delta}{v\sqrt{eH}\sin(\theta)}),\,\,\,\,\,\, B-phase
\eea
where $\kappa_{n}$ and $\kappa_{00}$ are the thermal conductivity in the normal
state and the thermal conductivity in the limit of universal heat conduction $\Gamma \rightarrow 0, T
\rightarrow 0.$  Here, $\Gamma$ is the quasiparticle scattering rate in the normal
state, and $\theta$ is the angle ${\bf H}$ makes from the $\hat{z}$ axis.
In both Eq.(5) and Eq.(7) we have assumed that the nodes in the B-phase are parallel
to the z axis.  Otherwise $\kappa_{zz}$ is smaller by a factor of $10 \sim 50$.  In
Fig. 4 we compare the observed angle dependent thermal conductivity with Eq.(6) and (7).
\begin{figure}[h]
\includegraphics[width=7cm]{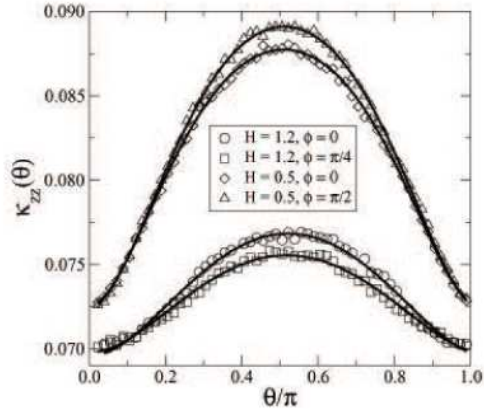}
\caption{Angular-dependent thermal conductivity in PrOs$_{4}$Sb$_{12}$.}
\end{figure}
These equations give an excellent fit.  From this we extract $v = 0.96 \times 10^{7}$ cm/sec
and $\Gamma$ = 0.1 K, where use is made of the weak-coupling theory gaps $\Delta_{A} = 4.2 K$ and
$\Delta_{B} = 3.5 K$ for the A and B phase respectively.  Note that de Haas-van Alphen measurements
\cite{44} give comparable values of v ($0.7\times 10^{7}$ cm/sec [$\alpha$-band], $0.6\times 10^{7}$ cm/sec
[$\beta$ - band] and $0.23 \times 10^{7}$ cm/sec [$\gamma$ band]).

\noindent{\it \bf 4. G-wave superconductivity in UPd$_2$Al$_3$}

This heavy-fermion superconductor with $T_{c} \simeq 2 K$ was discovered by Geibel et al
\cite{45} in 1991.  The reduction of the Knight shift in the superconducting state seen
in NMR \cite{46} and the Pauli limiting of H$_{c2}$ in UPd$_2$Al$_3$\cite{47} established spin-
singlet pairing.  Nodal superconductivity with horizontal nodes has been deduced from the
thermal conductivity \cite{48} and the c-axis tunneling data from UPd$_2$Al$_3$ thin films \cite{49}.
Here we focus on thermal conductivity data reported in Ref.\cite{13}.  First, $\kappa_{zz}$ at 
T = 0.4 K in a rotating magnetic field was measured.  For H $<$ 0.5 T no $\phi$ dependence was seen,
($\phi$ is the angle ${\bf H}$ makes from the x axis).  This indicates that the nodal lines
should be horizontal.  Second, $\kappa_{yy}$ in a magnetic field rotated within the z-x plane
was measured.

We show this in Fig. 5.  Following the standard procedure \cite{10}, the thermal 
\begin{figure}[h]
\includegraphics[width=6cm]{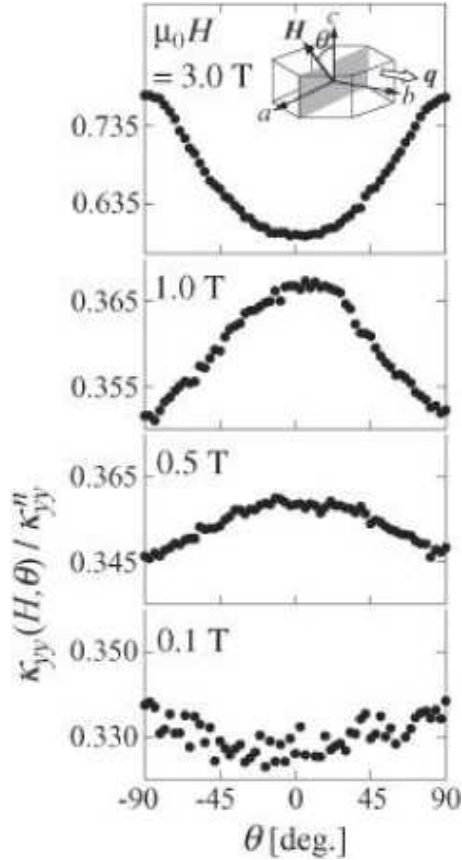}
\caption{Angular dependent magnetothermal conductivity in UPd$_{2}$Al$_{3}$}
\end{figure}
conductivity $\kappa_{yy}$ is obtained for a variety of $\Delta({\bf k})$ with
horizontal nodes as
\bea 
\kappa_{yy}/\kappa_{n} &=& \frac{2}{\pi} \frac{v_{a}^{2}eH}{\Delta^{2}} F_{1}(\theta)
\eea
in the superclean limit and 
\bea
\kappa_{yy}/\kappa_{00} &=& 1 + \frac{v_{a}^{2}eH}{6\pi \Gamma \Delta} F_{2}(\theta) 
\ln(2\sqrt{\frac{2\Delta}{\pi \Gamma}}) \ln(\frac{\Delta}{v\sqrt{eH}})
\eea
in the clean limit, where
\bea
F_{1}(\theta) &=& \sqrt{\cos^{2}\theta+\alpha \sin^{2}\theta} (1 + \sin^{2}\theta(-\frac{3}{8}
+ \alpha \sin^{2}\chi_{0}))\\
F_{2}(\theta) &=& \sqrt{\cos^{2}\theta+\alpha \sin^{2}\theta} (1 + \sin^{2}\theta(-\frac{1}{4}
+ \alpha \sin^{2}\chi_{0})),
\eea
and $\alpha = (v_{c}/v_{a})^{2}$. $\theta$ is the angle ${\bf H}$ makes from the z-axis and
$\chi_{0}$ is the nodal position.  For $\Delta({\bf k}) \sim \cos \chi, \cos(2 \chi),$ and
$\sin \chi$ we obtain $\chi_{0} = \frac{\pi}{2}, \frac{\pi}{4}$ and 0, respectively.  We
show $F_{1}(\theta)$ and $F_{2}(\theta)$ in Fig. 6 where we used $\alpha = 0.69$, the
appropriate value for UPd$_2$Al$_3$.  A comparison of Fig. 5 and Fig. 6 indicates
that $\Delta({\bf k}) \sim \cos(2 \chi)$ is the most appropriate choice.
\begin{figure}[h]
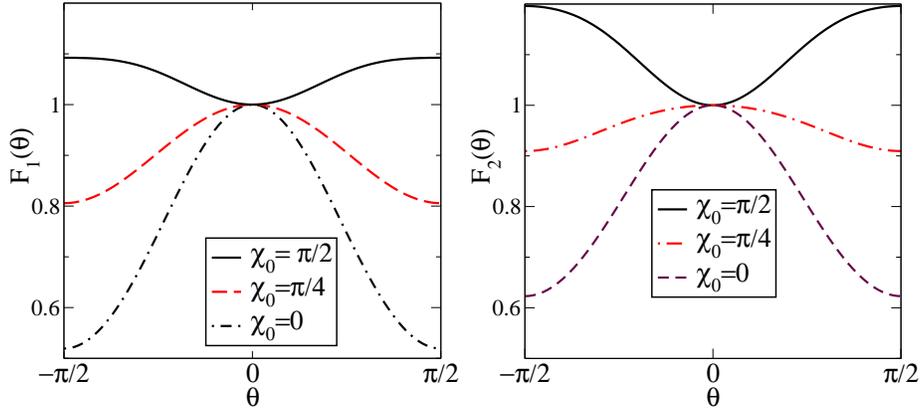

\includegraphics[width=6cm]{fig27a.eps}
\includegraphics[width=6cm]{fig27.eps}
\caption{Angular functions F$_{1}(\theta)$ (left) and F$_{2}(\theta)$ for various nodal positions}
\end{figure}
Similarly the universal heat conduction in nodal superconductors \cite{50,51} for a variety
of quasi-2D systems (f=$\cos(2\phi), \sin(2\phi), \cos \chi, cos(2\chi), \sin \chi$)
is a quantity of interest.  We obtain \cite{51}
\bea
\frac{\kappa_{xx}}{\kappa_{n}}&=& \frac{2 \Gamma_{0}}{\pi \Delta}\frac{1}{\sqrt{1+C_{0}^{2}}} 
E(\frac{1}{\sqrt{1+C_{0}^{2}}})= I_{1}(\frac{\Gamma}{\Gamma_{0}})
\eea
for all f's given above.  Here $\Gamma_{0}= 0.866 T_{c}$ and $C_{0}$ is determined by
\bea
\frac{C_{0}^{2}}{\sqrt{1+C_{0}^{2}}} K(\frac{1}{\sqrt{1+C_{0}^{2}}}) = \frac{\pi \Gamma}{2\Delta},
\eea
and K(z) and E(z) are the complete elliptic integrals.  Here $\kappa_{n}$ is the thermal
conductivity in the normal state with $\Gamma=\Gamma_{0}$.  Eq.(12) tells us that
$\kappa_{xx}$ cannot discriminate between different nodal structures.  On the other hand, we find
\bea
\frac{\kappa_{zz}}{\kappa_{n}}&=& I_{1}(\frac{\Gamma}{\Gamma_{0}}) \,\,\,for\,\,\, f=\cos(2\phi),\sin(2\phi) 
,\cos(2\chi),
\eea
but
\bea
\frac{\kappa_{zz}}{\kappa_{n}}&=& \frac{4\Gamma_{0}}{\pi \Delta \sqrt{1+C_{0}^{2}}}
(E(\frac{1}{\sqrt{1+C_{0}^{2}}})-C_{0}^{2} (K(\frac{1}{\sqrt{1+C_{0}^{2}}})-E(\frac{1}
{\sqrt{1+C_{0}^{2}}})))\\
&=& I_{2}(\frac{\Gamma}{\Gamma_{0}})\,\,\, for\,\,\, f=\cos\chi, e^{\pm i \phi} \cos\chi
\eea
and
\bea
\frac{\kappa_{zz}}{\kappa_{n}}&=& \frac{4\Gamma_{0} C_{0}^{2}}{\pi \Delta \sqrt{1+C_{0}^{2}}}
(K(\frac{1}{\sqrt{1+C_{0}^{2}}})-E(\frac{1}{\sqrt{1+C_{0}^{2}}}))\\
&=& I_{3}(\frac{\Gamma}{\Gamma_{0}})
\eea
for f=$\sin\chi$ and $e^{\pm i \phi} \sin \chi$.  We show $I_{1}(\Gamma/\Gamma_{0}), 
I_{2}(\Gamma/\Gamma_{0})$ and $I_{3}(\Gamma/\Gamma_{0})$ in Fig. 7.
\begin{figure}[h]
\includegraphics[width=7cm]{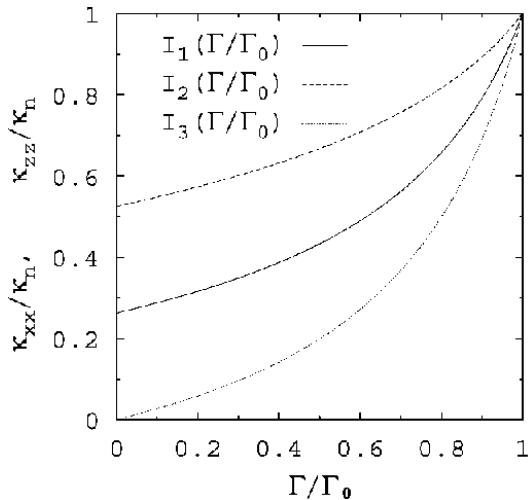}
\caption{The functions I$_{1}$, I$_{2}$ and I$_{3}$}
\end{figure}
Watanabe et al \cite{13} also measured $\kappa_{xx}$ and $\kappa_{zz}$ as a function of
${\bf H} ( || \hat{z})$.  Of course the effects of impurity scattering and of magnetic
fields are very different.  Nevertheless the comparison of these figures suggests again
$\Delta({\bf k}) \sim \cos(2\chi)$.

\noindent{\it \bf 5. Concluding Remarks}

We have surveyed recent developments on nodal superconductors.  As to the superconductivity
in Sr$_2$RuO$_4$ we believe the 2-D f-wave model with horizontal nodes is most promising,
in spite of the new specific heat data by Deguchi et al.  However, further 
experiments on Sr$_2$RuO$_4$ are clearly desirable.

The p+h-wave superconductivity in PrOs$_{4}$Sb$_{12}$ appears to solve many jigsaw puzzles
simultaneously.  These superconducting order parameters are highly degenerate due to multiple
chiral symmetry breaking.  We expect exciting topological defects in these systems associated
with the chiral symmetry breaking.

Also UPd$_2$Al$_3$ is the first Uranium compound examined through angle-dependent thermal 
conductivity experiments.  From a more indirect way, the gap symmetry of UPt$_3$ has been deduced to be
$f= e^{\pm 2i \phi}\sin^{2}\theta \cos \theta$ or E$_{2u}$ at least for the B phase\cite{52}.
It has been shown that there are many triplet superconductors, including UPt$_{3}$, UBe$_{13}$,
URu$_2$Si$_{2}$ and UNi$_{2}$Al$_{3}$ \cite{53}.  The determination of the gap symmetries of these
superconductors is of great interest.

Of course the gap symmetry itself cannot tell the underlying pairing mechanism of these 
systems.  But at least this provides the first important step for further exploration.  Also,
phonons most likely play no role in the pairing mechanism of most nodal superconductors.  The
majority of these pairings appear to be due to the antiparamagnon exchange.  But we can expect more
exotic interactions as well in this plethora of nodal superconductors.

{\bf Acknowledgments}

We have benefitted from helpful collaborations and discussions with Balazs Dora, Koichi Izawa,
Hae-Young Kee, Yuji Matsuda, Peter Thalmeier, Attila Virosztek and Tadataka Watanabe.  K.M. 
acknowledges gratefully the hospitality of the Max Planck Institute of the Physics of
Complex Systems at Dresden, where a part of this work was done.  S.H. was supported by the
NSF, Grant No. DMR-0089882.

\end{document}